\documentclass{emulateapj}
\usepackage{multirow}
\def\msun{$M_{\odot}$}

\shortauthors{Lin et al.}
\begin{document}

\title{The Megasecond {\it Chandra} X-Ray Visionary Project
  Observation of NGC 3115 (III): luminosity
  functions of LMXBs and dependence on stellar environments}

\author{Dacheng Lin\altaffilmark{1,2}, Jimmy A. Irwin\altaffilmark{1},
  Ka-wah Wong\altaffilmark{3}, Zachary G. Jennings\altaffilmark{4}, Jeroen Homan\altaffilmark{5}, Aaron
  J. Romanowsky\altaffilmark{4,6}, Jay Strader\altaffilmark{7}, Jean
  P. Brodie\altaffilmark{4},Gregory R. Sivakoff\altaffilmark{8}, Ronald A. Remillard\altaffilmark{5}}

\altaffiltext{1}{Department of Physics and Astronomy, University of
  Alabama, Box 870324, Tuscaloosa, AL 35487, USA}
\altaffiltext{2}{Space Science Center, University of New Hampshire, Durham, NH 03824, USA, email: dacheng.lin@unh.edu}
\altaffiltext{3}{Eureka Scientific, Inc., 2452 Delmer Street Suite 100, Oakland, CA 94602-3017}
\altaffiltext{4}{University of California Observatories, Santa Cruz, CA 95064, USA}
\altaffiltext{5}{MIT Kavli Institute for Astrophysics and Space Research, MIT, 70 Vassar Street, Cambridge, MA 02139-4307, USA}
\altaffiltext{6}{Department of Physics and Astronomy, San Jos\'{e} State University, One
Washington Square, San Jos\'{e}, CA 95192, USA}
\altaffiltext{7}{Department of Physics and Astronomy, Michigan State University,
East Lansing, Michigan, MI 48824, USA}
\altaffiltext{8}{Department of Physics, University of Alberta, Edmonton, Alberta,
T6G 2E1, Canada}

\begin{abstract}
 We have studied the X-ray luminosity function (XLF) of low-mass X-ray binaries (LMXBs) in the nearby lenticular galaxy NGC 3115, using the Megasecond \textit{Chandra} X-Ray Visionary Project Observation. With a total exposure time of $\sim$1.1 Ms, we constructed the XLF down to a limiting luminosity of $\sim$$10^{36}$ erg~s$^{-1}$, much deeper than typically reached for other early-type galaxies. We found significant flattening of the overall LMXB XLF from $dN/dL\propto L^{-2.2\pm0.4}$ above $5.5\times10^{37}$ erg~s$^{-1}$ to $dN/dL\propto L^{-1.0\pm0.1}$ below it, though we could not rule out a fit with a higher break at $\sim$$1.6\times10^{38}$ erg~s$^{-1}$. We also found evidence that the XLF of LMXBs in globular clusters (GCs) is overall flatter than that of field LMXBs. Thus our results for this galaxy do not support the idea that all LMXBs are formed in GCs. The XLF of field LMXBs seems to show spatial variation, with the XLF in the inner region of the galaxy being flatter than that in the outer region, probably due to contamination of LMXBs from undetected and/or disrupted GCs in the inner region. The XLF in the outer region is  probably the XLF of primordial field LMXBs, exhibiting $dN/dL\propto  L^{-1.2\pm0.1}$ up to a break close to the Eddington limit of neutron star LMXBs ($\sim$$1.7\times10^{38}$ erg~s$^{-1}$). The break of the GC LMXB XLF is lower, at $\sim$$1.1\times10^{37}$ erg~s$^{-1}$. We also confirm previous findings that the metal-rich/red GCs are more likely to host LMXBs than the metal-poor/blue GCs, which is more significant for more luminous LMXBs, and that more massive GCs are more likely to host LMXBs.
\end{abstract}


\section{INTRODUCTION}
Population studies of X-ray binaries in nearby galaxies have been made
possible thanks to the superb spatial resolution and excellent
sensitivity of the {\it Chandra X-ray Observatory}
\citep{wibrca2002}. The X-ray luminosity functions (XLFs) of point
sources have been obtained for many galaxies, and they are found to be
environment dependent \citep[see][for a review]{fa2006}. In young
normal galaxies, high-mass X-ray binaries (HMXBs) dominate, and the
XLFs follow a simple power law (PL) $dN/dL_{\rm X}\propto L_{\rm
  X}^{-\alpha}$ with $\alpha\approx 1.6$ over a large range of
luminosity: $10^{35}$ erg~s$^{-1}$ $\lesssim L_{\rm X}\lesssim10^{40}$
erg~s$^{-1}$ \citep{grgisu2003,migisu2012}. In old normal galaxies or
in the bulge of young normal galaxies, low-mass X-ray binaries (LMXBs)
dominate, and the XLFs seem relatively complicated, showing both a
high-luminosity break at $L_{\rm X}\sim 5\times 10^{38}$ erg s$^{-1}$
\citep{sairbr2001,gi2004,kifa2004,zhgibo2012} and a low-luminosity
break at $L_{\rm X}\sim 5\times 10^{37}$ erg s$^{-1}$
\citep{gi2004,vogisi2009,kifabr2009,zhgibo2012,lufafr2012}. The slope
between these two breaks is $\alpha\approx $1.8--2.2. Above the
high-luminosity break, the XLFs decrease sharply. Below the
low-luminosity break, the XLFs might flatten to $\alpha\approx 1.0$.

The high-luminosity break may be due to the Eddington limit of neutron
star (NS) LMXBs. The low-luminosity break has been attributed to
either the transition from mass transfer driven by magnetized stellar
wind at high luminosities to mass transfer driven by gravitational
wave emission at low luminosities \citep{poku2005}, or different types
of donor stars with the high-luminosity ones being giants and the
low-luminosity ones being main-sequence stars \citep{repoku2011}.  The
former predicts the slope below the break to be about 1.0 and the
slope above the break to be about 2.0. For the latter explanation, the
steepening of the XLF at high luminosity is due to the short life time
of binary systems with giants.

The XLFs of LMXBs have often been obtained by combining multiple
galaxies in order to improve the statistics. However, this method is
subject to the limitation that the normalizations of the XLFs in
different galaxies show a scatter of more than a factor of two
\citep{zhgibo2012}. The XLFs of LMXBs well below $10^{37}$ erg
s$^{-1}$ are still only obtained for very few old galaxies, most
notably Centaurus A and NGC 3379 \citep{vogisi2009,kifabr2009}, and
the bulge of M31 \citep{vogi2007a,vogi2007b}. A low detection limit is
critical for constraining the low-luminosity break, which has been
shown mostly for the old populations in young galaxies
\citep{fa2006}. Deep observations of old galaxies are also needed for
the investigation of the differences between the XLFs of LMXBs in
globular clusters (GCs) and in the stellar field, which can be used to
check whether they have the same origin. For instance, there is a
relative underabundance of faint LMXBs in GCs when compared with field
LMXBs \citep{vogisi2009,kifabr2009,zhgivo2011}.

\label{sec:intro}
\begin{figure} 
\centering
\includegraphics[width=3.4in]{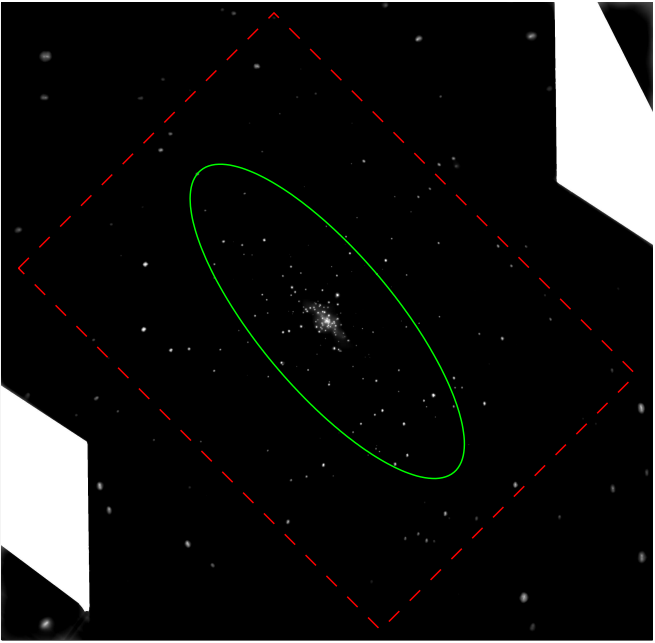}
\caption{{\it Chandra} X-ray image of NGC 3115 in 0.5--7 keV. The
  image is adaptively smoothed with the CIAO task csmooth and exposure
  corrected. The $D_{\rm 25}$ ellipse of the galaxy and the
  approximate FOV of the {\it HST}/ACS mosaic observation (dashed box)
  are also shown. \label{fig:colorimage}}
\end{figure}

\object{NGC 3115} was selected as the target of a 1 Megasecond {\it
  Chandra} X-ray Visionary Project (XVP) in Cycle 13. One main goal
was to study the gas flow inside the Bondi radius of the central
supermassive black hole (BH), which has been reported in
\citet{woirsh2014}.  The other goal was to have a deep look at the
X-ray binary population of a normal early-type galaxy. The detailed
analysis of the data, including the source list and detailed
properties of special sources, will be presented in
\citet[][Paper I hereafter]{liirwo2015a}. In the present paper, we concentrate on the XLF of
LMXBs, especially on its faint end below $10^{37}$ erg
s$^{-1}$. \object{NGC 3115} is a lenticular (S0) galaxy with an age of
$8.4\pm1.1$ Gyr \citep{sagoca2006} and at a distance of 9.7 Mpc
\citep{todrbl2001}. Including previous observations, the total
exposure time of {\it Chandra} on this galaxy is $\sim$1.1 Ms,
reaching a limiting luminosity of $\sim10^{36}$ erg s$^{-1}$. Thus it
is one of the best observed normal early-type galaxies by {\it
  Chandra}. Accompanying the {\it Chandra} XVP observation, there is a
six pointing {\it Hubble Space Telescope} ({\it HST}) mosaic
observation in the F475W and F850LP filters (hereafter $g$ and $z$
filters, respectively) using the Advanced Camera for Surveys (ACS). It
provides the information of GCs in the galaxy \citep{jestro2014},
which will be used by us to investigate the dependence of the XLF of
LMXBs on the stellar environment.

The paper is organized as follows. In Section~\ref{sec:reduction}, we
describe the X-ray data reduction, construction of XLFs,
incompleteness correction, and cross-correlation of the X-ray and
optical sources. In Section~\ref{sec:res}, we show the spatial
distributions of different populations, present the total LMXB XLF,
compare the XLFs of GC and field LMXBs, and investigate the GC LMXB
properties. In Section~\ref{sec:dis}, we discuss various caveats on
the XLFs that we obtain and the implication of our results for the
nature and formation of LMXBs. Our conclusions are given in
Section~\ref{sec:con}.

\section{DATA ANALYSIS}
\label{sec:reduction}

\subsection{Observations and Source Detection}

\tabletypesize{\scriptsize}
\begin{deluxetable}{rrccc}
\tablecaption{Observation Log\label{tbl:obslog}}
\tablewidth{0pt}
\tablehead{\colhead{Notation} & \colhead{Obs. ID} &\colhead{Date} &\colhead{Exposure} &\colhead{Offset\tablenotemark{a}}\\
& &  & (ks) & (arcmin)}
\startdata
1 & 2040 & 2001-06-14 & 35.8 & 1.5\\
2 & 11268 & 2010-01-27 & 40.6 & 0.1\\
3 & 12095 & 2010-01-29 & 75.6 & 0.1\\
4 & 13817 & 2012-01-18 & 171.9 & 0.0\\
5 & 13822 & 2012-01-21 & 156.6 & 0.0\\
6 & 13819 & 2012-01-26 & 72.9 & 0.0\\
7 & 13820 & 2012-01-31 & 184.1 & 0.0\\
8 & 13821 & 2012-02-03 & 157.9 & 0.0\\
9 & 14383 & 2012-04-04 & 119.4 & 0.3\\
10 & 14419 & 2012-04-05 & 46.3 & 0.3\\
11 & 14384 & 2012-04-06 & 69.7 & 0.3
\enddata 
\tablenotetext{a}{Aim point offset from observation 13820.}
\end{deluxetable}

The {\it Chandra} observations of NGC 3115 are listed in
Table~\ref{tbl:obslog}. There are 11 observations in total from
essentially three epochs: one in 2001, two in 2010 and nine in 2012.
All observations used the imaging array of the AXAF CCD Imaging
Spectrometer \citep[ACIS; ][]{bapiba1998}. The reduction of the data
and the creation of the source list were presented in detail in Paper
I, and here we briefly summarize the procedure adopted. The data were
analyzed with the {\it Chandra} Interactive Analysis of Observations
(CIAO, version 4.6) package.  The data were reprocessed to apply the
latest calibration (CALDB 4.5.9) and the subpixel algorithm
\citep{likapr2004} using the CIAO script {\it chandra\_repro}. Some
short background flares seen in observations 2040, 13819 and 13822
were excluded. The final exposure used for each observation is given
in Table~\ref{tbl:obslog}. The relative astrometry between
observations was corrected, and the source detection was performed on
individual observations as well as on the merged one using the 0.5--7
keV energy band with the CIAO {\it wavdetect} wavelet-based source
detection algorithm \citep{frkaro2002}.  We used two different image
binning resolutions: one at single sky pixel resolution (0\farcs492)
over the full field of view (FOV) and the other at 1$/$8 sky pixel
resolution covering an area of 3$\arcmin$$\times$3$\arcmin$ centered
at the center of NGC 3115. The subpixel binning images were used to
improve the spatial resolution of the crowded field near the center of
the galaxy. Sources detected from the merged observation and
individual observations were cross-correlated to create the final
unique source list.

For each unique source, we extracted the source and background spectra
and created the response file for each individual observation. They
were then merged to create the spectra and response files for the
merged observation. The source region was set to be a circle enclosing
90\% of the point spread function (PSF) at 2.3 keV. The background
region was set to be a concentric annulus, with inner and outer radii
of two and five times the source radius, respectively. The
background-subtracted 0.5--7 keV count rates were then converted to
unabsorbed fluxes and luminosities, with the conversion factors based
on the corresponding response files and assuming an absorbed PL
spectral shape with a photon index of 1.7 and the Galactic
absorption $N_{\rm H}=4.32\times10^{20}$ cm$^{-2}$ \citep{kabuha2005}.

\subsection{Incompleteness Calculation and XLF Construction}
The point-source detection sensitivity varies across the {\it Chandra}
image, owing to the position dependence of the diffuse X-ray emission
in the galaxy, the PSF extent, the exposure, and CCD
efficiency. Therefore it is necessary to carry out the incompleteness
correction for the XLF. The $D_{\rm 25}$ region of NGC 3115 has a
semi-major axis of $a=3.62\arcmin$ (10.2 kpc), a semi-minor axis of
$b=1.23\arcmin$ (3.5 kpc) and a position angle of $40\degr$
\citep{dedeco1991}.  To limit the incompleteness effects and the
cosmic X-ray background (CXB) contribution, we defined our study field
for the XLF of field LMXBs as the region inside $D_{\rm 25}$. Further
considering that the central region is very crowded and has strong
diffuse X-ray emission, we excluded the central $a=10\arcsec$
elliptical region (the eccentricity and position angle follow the
$D_{\rm 25}$ ellipse) for all XLFs throughout the paper.

We calculated the incompleteness function $K(L)$, the fraction of
pixels weighted by the assumed spatial distribution of sources, in
which a source with the luminosity $L$ or higher would be detected,
using the backward correction method \citep{kifa2003}. In this method,
sources are simulated with MARX and added one by one to the real
observed image, which is then checked to see whether each one could be
detected with {\it wavdetect}. The source spectral shape was assumed to be a
PL with a photon index of 1.7 and the Galactic absorption. The
simulations were carried out for a series of luminosities with an
increasing factor of 1.1 and 1.21 below and above the 90\% completeness
luminosity, respectively. The positions of simulated sources for each
luminosity were specified as follows. The $D_{\rm 25}$ ellipse was
divided into elliptical annuli with a series of ellipses that have the
eccentricity and the position angle following $D_{\rm 25}$ and
semi-major axis $a$ for the $i$th ($i=0$ to 56) ellipse assuming
$a_i=a_{i-1}+{\delta a}*1.04^{i-1}$, where $a_0=10\arcsec$ and $\delta
a=1\arcsec$. Each elliptical annulus was then divided into 80 cells
with an equal area, and the simulated source position was specified at
the center of each cell, with 4560 in total. In some cases the
simulated sources coincided with the real sources, and we assumed that
the simulated sources were detected by {\it wavdetect} only if the simulated
sources dominate the flux over the real sources, which is to take into
account the source confusion effect. We calculated $K(L)$ for the CXB
sources and field LMXBs separately, because the CXB sources have a
flat distribution and the field LMXBs are expected to follow the
$K_{\rm S}$-band light \citep{gi2004}, for which we used the 2MASS
Large Galaxy Atlas data \citep{jachcu2003}.

We also calculated $K(L)$ for GC LMXBs. We used all the GCs detected
in the optical (Section~\ref{sec:crcorrel}) as the parent spatial
distribution of GC LMXBs and assumed that they have equal probability
of hosting an LMXB. To limit the incompleteness effects and spurious
rate of the GC LMXB identification, our study field for the XLF of GC
LMXBs is set to be the {\it HST}/ACS field of view (FOV). The {\it
  HST}/ACS FOV reached $\sim$1.3$D_{\rm 25}$ and $\sim$2.5$D_{\rm
  25}$ in the major-axis and minor-axis directions of the $D_{\rm 25}$
ellipse, respectively (Figure~\ref{fig:colorimage}).

The differential XLF of LMXBs in a given region can be calculated as
follows \citep[refer to, e.g.,][]{vogisi2009}:
\begin{equation}
\frac{dN_{\rm LMXB}}{dL}=\frac{1}{K_{\rm LMXB}(L)}\left(\frac{dN_{\rm obs}}{dL}-{4\pi D^2}K_{\rm
    CXB}(L)\frac{dN_{\rm CXB}}{dL}\right),
\label{eq:xlfdef}
\end{equation}
where $N_{\rm obs}$ is the total number of observed sources and $D$ is
the distance to NGC 3115. The quantity $4\pi D^2dN_{\rm CXB}/{dL}$ is
$dN_{\rm CXB}/{dS}$, the $\log(N)$--$\log(S)$ distribution of the CXB
sources. We used the full band (0.5--10 keV) $\log(N)$--$\log(S)$
distribution of CXB sources from \citet{genala2008}, with their
0.5--10 keV flux converted to our 0.5--7 keV band assuming a PL
spectrum with a photon index of 1.4. For the XLF of field LMXBs only,
we filtered out GC LMXBs and had the CXB contribution estimated as
described above. For the XLF of GC LMXBs, the CXB contribution was not
corrected because it is negligibly small.

We did not correct XLFs for the HMXB contribution. Following
\citet{migisu2012}, we estimated the star formation rate in NGC 3115
to be 0.07 \msun\ yr$^{-1}$ (see their Equation 9, which is based on
the UV and IR emission). Based on their XLF for HMXBs (their Equation
18), we can estimate the number of HMXBs in NGC 3115 above $L_{\rm
  lim}$ to be 2.3, which is one order of magnitude less than the CXB
contribution and is thus negligibly small.

To compare with previous studies, some XLFs presented in this
  study will be divided by (thus normalized to) the stellar mass
enclosed in our study region of the field LMXB XLF (i.e., within
$D_{\rm 25}$ and outside the central $a=10\arcsec$ ellipse). Following
\citet{zhgibo2012}, we used the $K_{\rm s}$-band luminosity and
estimated the stellar mass in our study region to be
$6.31\times10^{10}$ \msun\ (the total stellar mass within $D_{\rm 25}$
is $7.83\times10^{10}$ \msun).

\subsection{Multiwavelength cross-correlation}
\label{sec:crcorrel}
We cross-correlated our X-ray sources with optical sources from {\it
  HST}/ACS mosaic imaging and Subaru/Suprime-Cam imaging to search for
the GC LMXBs. \citet{jestro2014} compiled 360 GC candidates from {\it
  HST}/ACS mosaic imaging and an additional 421 from
Subaru/Suprime-Cam imaging \citep{arrobr2011}. Before
cross-correlation, we first carried out absolute astrometry correction
on X-ray sources by cross-correlating their positions with the 360 GC
candidates from {\it HST}/ACS mosaic imaging, whose astrometry was
registered to the USNO-B1.0 Catalog \citep{moleca2003}. We only used
X-ray sources detected at $>6\sigma$ significance and with off-axis
angles $<6\arcmin$ (the limit of {\it HST}/ACS FOV) in the
cross-correlation. We found 30 matches with a median separation
residual of 0.06$\arcsec$.

We searched for the {\it HST}/ACS GC counterparts to our X-ray sources
using the 99.73\% (i.e., 3$\sigma$) positional uncertainty that combines
both X-ray and optical components. For the {\it HST}/ACS sources, we
assumed the half light radius as the 1-$\sigma$ positional uncertainty. We
also included a systematic uncertainty which was assumed to be 0.05$\arcsec$
(1$\sigma$, in both R.A. and Decl.) based on the above matches in the
absolute astrometry correction. This systematic uncertainty is probably
overestimated, but it is so small that the number of GC matches
remained the same even if we did not include this systematic
uncertainty. The offsets of all matches (37 in total) are $<1\arcsec$ (only
3 have offsets $>0.2\arcsec$) and have a median of 0.07$\arcsec$. To
estimate the spurious rate, we rotated the {\it HST}/ACS field by
$\pm$10$\degr$, 180$\degr$$\pm$10$\degr$, and 180$\degr$ around the
galaxy center and carried out the cross-correlation in the same way
and found the spurious rate to be about 3\%.

The {\it HST}/ACS GC distribution from \citet{jestro2014} decreases
sharply within $\sim$$0.25D_{\rm 25}$ (see
Section~\ref{sec:spatialdis}). Therefore their GC list is probably
fairly incomplete in this region owing to strong stellar light. We
tried to match our X-ray sources with the sources that were detected
by \citet{jestro2014} but not classified as GCs, and we found three
extra matches (S12, S53 and S79 in Paper I) within $0.25D_{\rm
  25}$. One more source (S65) seems to have an optical match from our
visual inspection but it is not detected by \citet{jestro2014} due to
its being too close to the galaxy center (5.5$\arcsec$, within the
$a=10\arcsec$ elliptical exclusion region). We classify these four
sources as GC LMXB candidates. They are all bright ($>$10$^{37}$
erg~s$^{-1}$) and are not expected to be CXB sources (the expected CXB
source number at this luminosity is 0.5 and is much smaller if only
CXB sources with bright optical counterparts are considered). Another
source (S92 in Paper I) outside $0.25D_{\rm 25}$ also has an optical
match, but it was classified as a star by \citet{jestro2014} due to
the measurement of a radial velocity (238 km s$^{-1}$) much lower than
the threshold of 350 km s$^{-1}$ that they adopted to define
GCs. Considering that the size and color of this optical match are
consistent with typical GCs, we treat it as a GC LMXB candidate
too. We did not include the above five LMXBs in either field or GC
XLFs (although S65 is outside the study region and would not be
included anyway).

In our search for the GC optical counterparts to our sources detected
only in the Subaru/Suprime-Cam imaging (i.e., not in the {\it HST}/ACS
imaging), we also used the 99.73\% positional uncertainty. The 1-$\sigma$
positional uncertainty of the optical sources was assumed to be
0.1$\arcsec$ in both R.A. and Decl. The 1-$\sigma$ systematic uncertainty
was assumed to be 0.1$\arcsec$ in both R.A. and Decl. (Paper I). To
limit the spurious rate, we have a maximum searching radius of
2$\arcsec$. The spurious rate was estimated to be 5\% (Paper I). Only
eight Subaru/Suprime-Cam GC matches were found. They are used only for
the study of the spatial distribution of GC LMXBs in
Section~\ref{sec:spatialdis} but not for the study of XLFs.

\section{RESULTS}
\label{sec:res}

\begin{figure} 
\centering
\includegraphics[width=3.4in]{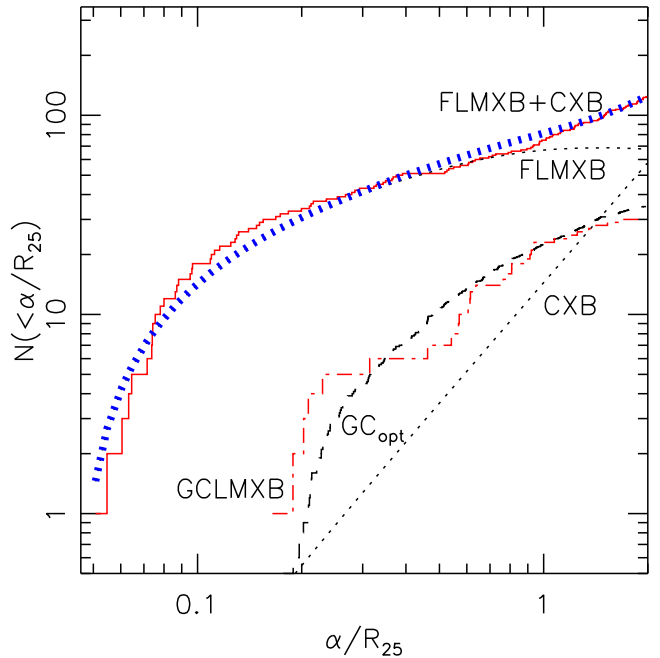}
\caption{The radial distribution of observed sources excluding GC
  LMXBs (red solid line), compared to the model (thick blue dotted
  line) composed of field LMXBs (FLMXB) and CXB sources (thin dotted
  line). Also plotted are the radial distributions of GC LMXBs (red
  dot-dashed line) and the optical GCs (dashed line, divided by a
  factor of 10). The cumulation starts at $\alpha/R_{\rm 25}=0.046$
  (i.e., excluding the central $a=10\arcsec$ elliptical region) and
  ends at $2D_\mathrm{25}$. To reduce the incompleteness effects, only sources
  with $L_{\rm X}\ge4.0\times10^{36}$ erg s$^{-1}$ are
  used. \label{fig:spatialdist}}
\end{figure}

\begin{figure} 
\centering
\includegraphics[width=3.4in]{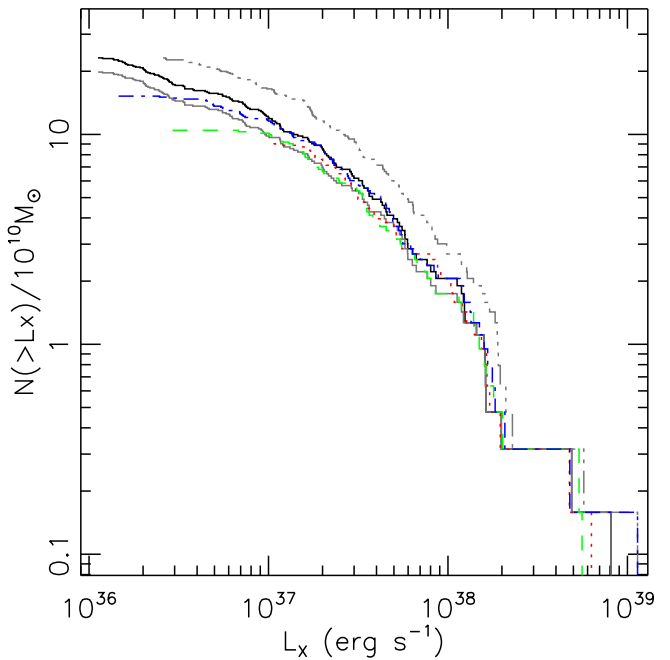}
\caption{Cumulative XLFs from the merged observation (black solid
  line) and single observations 2040 (red dotted line), 12095 (green
  dashed line), and 13820 (blue dot-dashed line). The grey solid line
  is also from the merged observation but using sources with $V_{\rm
    var}<5.0$. The grey dot-dot-dashed line uses the maximum
  luminosity of each source. The distributions are not corrected for
  incompleteness or the CXB contribution. \label{fig:lumdis_ind} }
\end{figure}

\begin{figure} 
\centering
\includegraphics[width=3.4in]{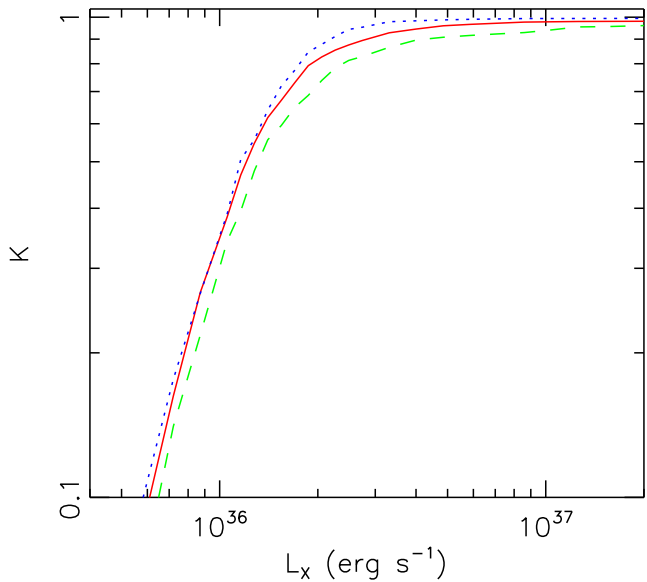}
\caption{The incompleteness functions for field LMXBs (red solid
  line), GC LMXBs (green dashed line) and CXB sources (blue dotted line). \label{fig:incompletefun}}
\end{figure}

\begin{figure*} 
\centering
\includegraphics[width=6.8in]{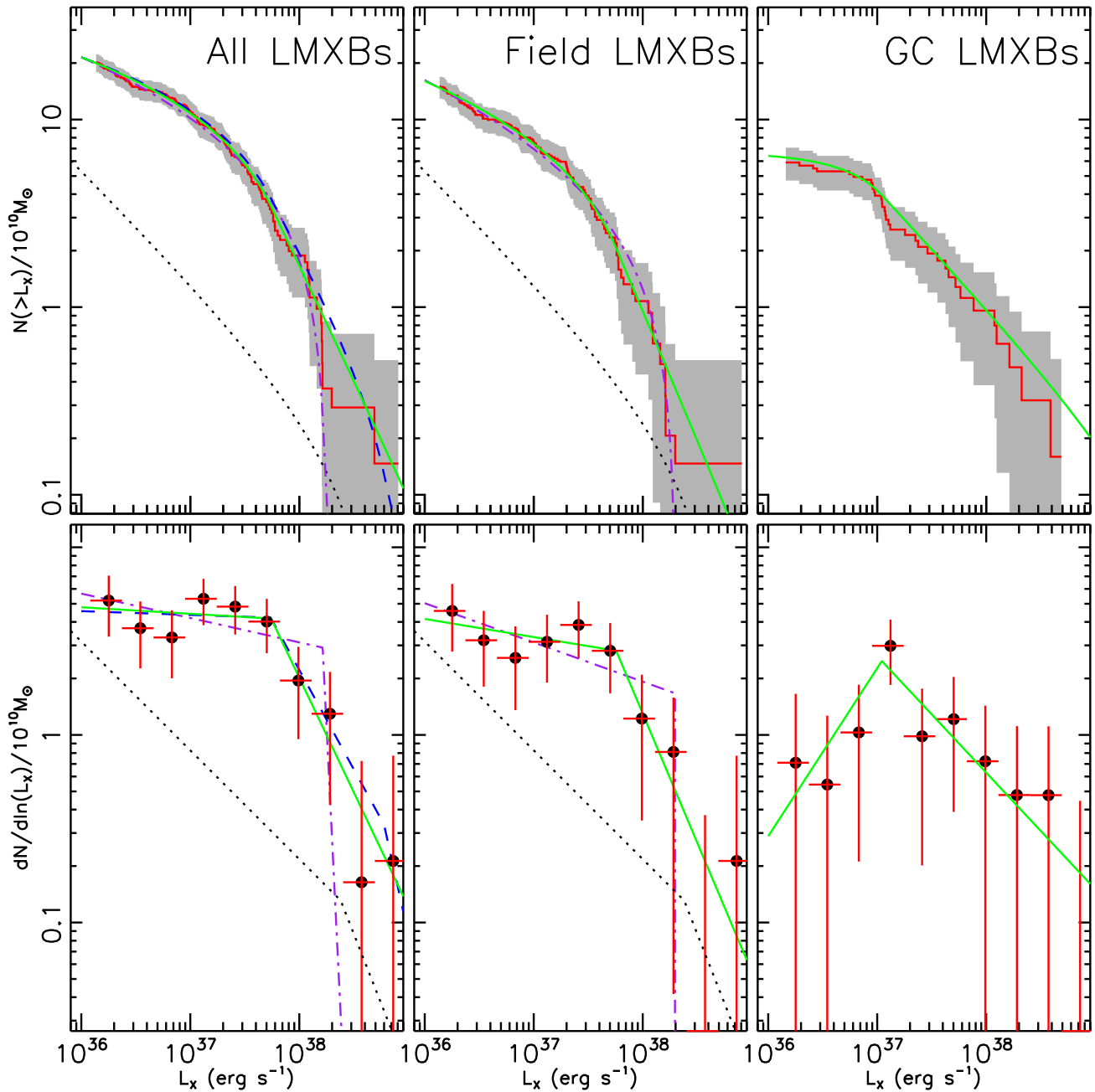}
\caption{Incompleteness corrected and CXB contribution subtracted XLFs
  for all LMXBs (left panels), field LMXBs (middle panels) and GC
  LMXBs (right panels). We plot both the cumulative (upper panels,
  the red solid line with the shaded area representing $1\sigma$
  Poissonian uncertainty) and differential forms (lower panels). Our
  fits with a broken PL are shown as green solid lines (the low-break
  solution) and purple dot-dashed lines (the high-break solution, when
  present, see text). We note that  the
    high-break solutions have a large uncertainty in the second index,
    whose lower limit is given in Table~\ref{tbl:xlffit}. The dashed blue lines in
  the left panels are the fit to the average XLF of 20 early-type
  galaxies by \citet{zhgibo2012}, with the normalization decreased by
  24\%. The black dotted line is the expected CXB
  distribution. \label{fig:lumdis_sum_subcxb}}
\end{figure*}
 
\begin{figure} 
\centering
\includegraphics[width=3.4in]{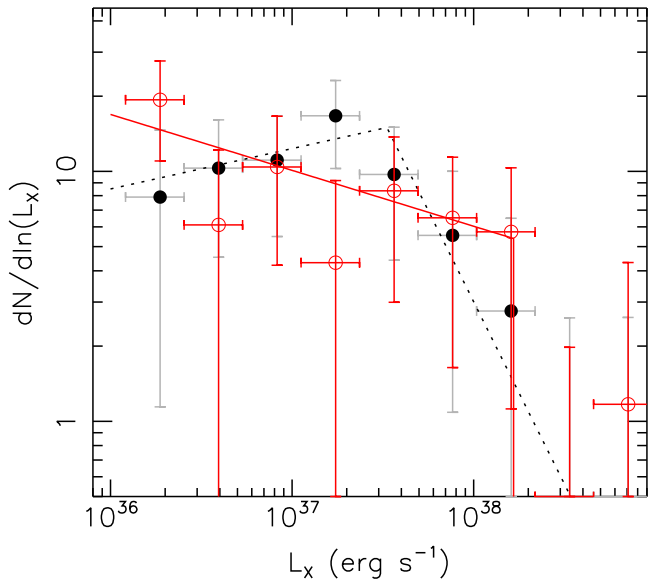}
\caption{Incompleteness corrected and CXB contribution subtracted XLFs of field LMXBs within $\alpha/R_{\rm
    25}=$ 0.046--0.2 (filled circles) and those within  $\alpha/R_{\rm
    25}=$ 0.2--1.0 (red open circles). Their best-fitting broken PL
  is shown as a dotted line and a red solid line, respectively.\label{fig:fieldlmxbxlf_sub}}
\end{figure}

\begin{figure} 
\centering
\includegraphics[width=3.4in]{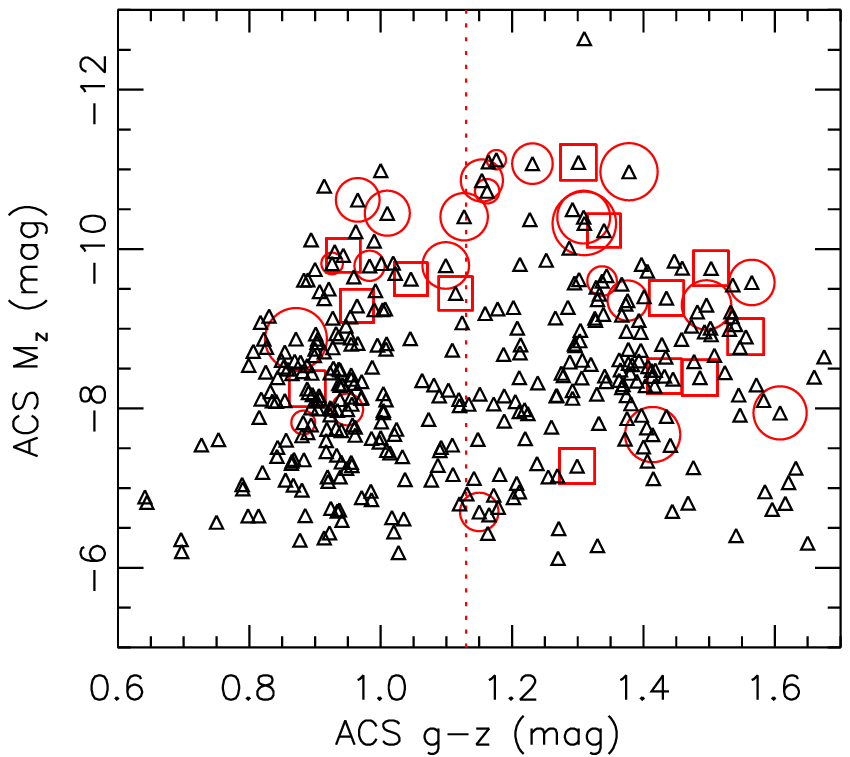}
\caption{The color-magnitude diagram of {\it HST}/ACS GCs, with those
  containing LMXBs enclosed with squares ($L_{\rm X}$ within
  (0.8--1.3)$\times10^{37}$ erg s$^{-1}$) or circles (others), whose
  size is proportional to the logarithm of the LMXB luminosity. The
  red dotted line at $g-z=1.13$ mag was used by \citet{jestro2014} as
  the dividing line between blue/metal-poor and the red/metal-rich
  subpopulations. \label{fig:gcpro}}
\end{figure}

\subsection{Spatial Distribution}
\label{sec:spatialdis}
Figure~\ref{fig:colorimage} shows the {\it Chandra} X-ray image of NGC
3115. From the merged observation, we detected 145 X-ray sources above
the 50\% completeness luminosity $L_{\rm lim}=1.21\times10^{36}$ erg
s$^{-1}$ (Section~\ref{sec:xlfs}) within $D_{\rm 25}$ and outside the
central $a=10\arcsec$ ellipse (23 within this exclusion region).  The
expected CXB source number is 26.5.

Figure~\ref{fig:spatialdist} shows the cumulative radial distributions
of different classes of objects detected from the merged
observation. Due to the high inclination
\citep[$i=86\degr$,][]{caheni1987} of NGC 3115, we plot the
distributions with respect to $\alpha/R_{\rm 25}$, where $\alpha$ is
the angular separation between the source and the galaxy center and
$R_{\rm 25}$ is the elliptical radius of the $D_{25}$ isophotal
ellipse in the direction from the galaxy center to the source.  The
cumulation starts at $\alpha/R_{\rm 25}=0.046$ because we have
excluded the central $a=10\arcsec$ elliptical region in the
calculation of XLFs. To reduce the incompleteness effects, we used only
sources above $4\times10^{36}$ erg s$^{-1}$, which is the 82\% completeness
luminosity for CXB sources and the 94\% completeness luminosity for field
LMXBs within $2D_\mathrm{25}$ (but outside the central $a=10\arcsec$ elliptical region). The non-GC X-ray
sources, expected to consist of field LMXBs and CXB sources mostly
(red solid line), and GC LMXBs (red dot-dashed line) are plotted
separately. The thick blue dotted line models the spatial distribution
of non-GC X-ray sources using two components, one for CXB sources and
the other for field LMXBs, which was assumed to follow the IR light in
the $K_{\rm s}$ band. The normalization of the field LMXB component
was determined so as to give the same number of sources, after adding
the CXB component, as observed within $2D_{\rm 25}$. The observed
distribution roughly follows this model, with the Kolmogorov-Smirnov
(K-S) test giving a probability of 30\%.  The radial distribution of
GC LMXBs seems to approximately follow that of GCs detected in the
optical as well, with the K-S test giving a probability of 62\%.

Figure~\ref{fig:spatialdist} shows a dramatic difference between the
spatial distributions of field and GC LMXBs. The field LMXBs tend to
cluster toward the galaxy center, while GC LMXBs tend to be more
spread out. Only 7 out of 360 {\it HST}/ACS GCs are within $0.2D_{\rm
  25}$. However, we note that the GC detection is most probably fairly
incomplete near the galaxy center due to strong stellar light. Some
GCs could also be destructed near the galaxy center due to mass
segregation.

There are 50 non-GC X-ray sources above $4\times10^{36}$ erg s$^{-1}$
observed between $D_{25}$ and $2D_{25}$, while the expected number of
CXB sources is 40.2 after incompleteness correction and the expected
number of field LMXBs are 1.5 based on the IR light in the $K_{\rm s}$
band. The above 50 sources include two special sources: S109 and S179
(Paper I). The former is a supersoft X-ray source (SSS) at
1.53$D_{25}$. The latter is a transient with 0.5--7 keV long-term
variability factor $V_{\rm var}>20$ and relatively soft X-ray spectra
(classified as a BH X-ray binary candidate in Paper I), and it is at
1.04$D_{25}$. Therefore these two sources are most probably in NGC
3115, instead of being CXB sources. Then we have 48 left, which is
19\% more than the expected number of CXB sources.  If we concentrate
on the region between $1.3D_{25}$ and $2D_{25}$ (there is little IR
light outside $1.3D_{25}$ from the galaxy), we have 33 non-GC X-ray
sources above $4\times10^{36}$ erg s$^{-1}$, excluding S109, and this
number is very close to the expected number of CXB sources (30.6,
incompleteness corrected).

\subsection{The XLF of All LMXBs}
\label{sec:xlfs}
\tabletypesize{\scriptsize}
\setlength{\tabcolsep}{0.06in}
\begin{deluxetable}{lccccccc}
\tablecaption{The maximum likelihood fits to XLFs of different
  populations using a broken powerlaw \label{tbl:xlffit}}
\tablewidth{0pt}
\tablehead{\colhead{Population} &\colhead{$N_{\rm tot}$} &\colhead{$N_{\rm CXB}$} &\colhead{$\alpha_1$}  &\colhead{$\alpha_2$} &\colhead{$L_{\rm b}$} &\colhead{$K$}\\
(1)&(2) & (3) &(4) & (5) &(6) & (7)}
\startdata
\multirow{2}{*}{Total} &\multirow{2}{*}{145} & \multirow{2}{*}{26.5} & $1.03_{-0.14}^{+0.10}$&$2.2_{-0.4}^{+0.3}$&$55_{-23}^{+12}$&$30_{-9}^{+9}$\\
 & &  & $1.13_{-0.08}^{+0.08}$&$>6.7$&$164_{-8}^{+13}$&$36_{-7}^{+9}$\\ 
\cline{1-7}
\multirow{2}{*}{Field} & \multirow{2}{*}{114} & \multirow{2}{*}{26.5}&$1.10_{-0.13}^{+0.12}$ & $2.4_{-0.5}^{+0.5}$ &$57_{-24}^{+17}$ & $23_{-6}^{+8}$\\
 &  & &$1.21_{-0.09}^{+0.09}$ & $>7.0$ &$198_{-13}^{+11}$ & $32_{-7}^{+9}$\\
Field$_{\rm in}$ & 44 &1.0 & $0.84_{-0.34}^{+0.22}$ & $2.5_{-0.6}^{+1.1}$ &$33_{-15}^{+27}$ & $9_{-4}^{+5}$\\
Field$_{\rm out}$ & 70 & 25.5& $1.20_{-0.15}^{+0.14}$ & $>9.6$ &$166_{-6}^{+17}$ & $15_{-5}^{+7}$\\
\cline{1-7}
GC$_{\rm ACS}$ &36&0 &$0.10_{-0.62}^{+0.39}$&$1.6_{-0.2}^{+0.2}$&$11_{-3}^{+2}$&$1.8_{-1.2}^{+2.1}$\\
GC$_{\rm D25}$ & 27 & 0 & $0.41_{-0.56}^{+0.39}$   &$1.6_{-0.2}^{+0.2}$&$12_{-3}^{+34}$&$2.5_{-1.4}^{+2.8}$
\enddata 
\tablecomments{Columns are as follows. (1) The population, (2) the
  total number of sources observed above $L_{\rm
    lim}=1.21\times10^{36}$ erg~s$^{-1}$, (3) the expected observed
  number of CXB sources above $L_{\rm lim}$ based on
  \citet{genala2008}, (4) initial slope, (5) second slope, (6) break
  luminosity in units of $10^{36}$ erg~s$^{-1}$, (7) the normalization
  (not normalized by the stellar mass).  All errors and lower limits
  are at the $1\sigma$ level. Population descriptors: ``total'': all
  LMXBs within (0.046--1.0)$D_{\rm 25}$, ``field'': all LMXBs within
  (0.046--1.0)$D_{\rm 25}$ but excluding all 27 GC LMXBs and four GC
  LMXB candidates in the region, ``field$_{\rm in}$'': similar to
  ``field'' but only within (0.046--0.2)$D_{\rm 25}$, ``field$_{\rm
    out}$'': similar to ``field'' but only within (0.2--1.0)$D_{\rm
    25}$, GC$_{\rm ACS}$: all {\it HST}/ACS GC LMXBs, GC$_{\rm D25}$:
  all {\it HST}/ACS GC LMXBs within $D_{\rm 25}$.  For the ``total''
  and ``field'' XLFs, we give both the low-break and high-break
  solutions. }
\end{deluxetable}

Because a large fraction of our sources are variable, we first check
whether there is any difference in XLFs between observations. We plot
the observed cumulative XLFs for the merged observation and the
longest observation in each of the three epochs in
Figure~\ref{fig:lumdis_ind}. We compared their XLFs using the K-S
test. We focused on luminosities above the 90\% completeness limit,
which are $3.3\times10^{37}$ erg s$^{-1}$, $1.6\times10^{37}$ erg
s$^{-1}$, and $8.6\times10^{36}$ erg s$^{-1}$ for observations 2040,
12095 and 13820, respectively. We found that the XLFs obtained in
these individual observations are consistent with the same
distribution as that obtained in the merged observation with the K-S
probability of 48\%, 98\%, and 82\%, respectively. \citet{vogisi2009}
obtained a similar conclusion for Centaurus A. We also constructed the
XLF using less variable sources (the 0.5--7 keV flux long-term
variability $V_{\rm var}<5$, Paper 1) in the merged observation (gray
solid line in Figure~\ref{fig:lumdis_ind}) and compared it with the
XLF using all sources. Above the luminosity at the 90\% completeness
level ($2.7\times10^{36}$ erg s$^{-1}$), the K-S test also indicates
no obvious difference between them, with the probability of
100\%. Finally, we constructed the XLF using the maximum luminosity of
each source (gray dot-dot-dashed line in
Figure~\ref{fig:lumdis_ind}). When we compared it with the XLF from
the merged observation and limited to $L>2.7\times10^{36}$ erg
s$^{-1}$, the K-S test gave a probability of 37\%. From all the above
comparison, we see no significant effect of the source long-term
variability on the XLFs. Therefore, we will use the mean luminosities
of the sources detected from the merged observations for the XLFs
hereafter.

The incompleteness functions for (field) LMXBs and CXB sources are
shown in Figure~\ref{fig:incompletefun}. While they appear to be
similar to each other, this is a coincidence, as they can be very
different if different regions other than the whole $D_{25}$
(excluding the central $a=10\arcsec$ ellipse) are used. The luminosity
corresponding to the 50\% completeness level is about $L_{\rm
  lim}=1.21\times10^{36}$ erg s$^{-1}$, which is much lower than
typical values of $\sim10^{37}$ erg s$^{-1}$ achieved for other
galaxies by {\it Chandra} \citep[e.g.,][]{zhgibo2012}. Above this
luminosity limit, the sources detected in the merged observation
within $D_{\rm 25}$ are all $>$2.9$\sigma$.

The incompleteness corrected and CXB contribution subtracted XLF above
$L_{\rm lim}$ for all LMXBs is plotted in the left panels in
Figure~\ref{fig:lumdis_sum_subcxb}.  The XLF is steep down to a break
around $5\times10^{37}$ erg s$^{-1}$, below which the XLF flattens
clearly. We fitted the differential form of XLF with a small $L_{\rm
  X}$ bin size of $\delta\log(L_{\rm X})=0.02$ using the C statistic
(it is maximum likelihood-based) in the Xspec fitting package
\citep{ar1996}. In this way, the observed XLF was used in the fit, but
the fitting model was modified by the incompleteness function (through
a response file).  The model that we adopted is a broken PL:
\begin{eqnarray}
\frac{dN}{dL_{36}}=\left\{ \begin{array}{ll}
\renewcommand{\arraystretch}{3}
K L_{36} ^{-\alpha_1},
                        & \mbox{\hspace{1.5cm} $L_{36}<L_{\rm b}$}\\
K L_{\rm b}^{\alpha_2-\alpha_1}L_{36}^{-\alpha_2},
                        & \mbox{\hspace{1.5cm} $L_{36} >L_{\rm b}$}\\
\end{array}
, \right.
\label{eq:uxlf}
\end{eqnarray}
where $L_{36}=L_{\rm X}/(10^{36}\ {\rm erg\ s}^{-1})$. We obtained
$\alpha_1=1.03_{-0.14}^{+0.10}$, $\alpha_2=2.2^{+0.3}_{-0.4}$, and
$L_{\rm b}=55^{+12}_{-23}$ (Table~\ref{tbl:xlffit} and
Figure~\ref{fig:lumdis_sum_subcxb} (green solid
line)).

\citet{zhgibo2012} fitted the average XLF of 20 early-type galaxies
with the template introduced by \citet{gi2004}, which is essentially a
double broken PL. Because we do not have enough statistics above the
second break that they obtained ($6\times10^{38}$ erg s$^{-1}$), we do
not need to introduce the second break to fit our XLF. Our fit is
fully consistent with that obtained by \citet{zhgibo2012}, who
reported $\alpha_1=1.02\pm0.08$, $\alpha_2=2.06\pm0.06$, and $L_{\rm
  b}=54.6\pm4.0$. The main difference is the normalization $K$, with
ours being about 76\% of that of \citet{zhgibo2012}, after being
normalized by stellar mass (our $K=4.6\pm1.3$ per $10^{10}$
M$_{\odot}$ versus their $K=6.0\pm1.7$ per $10^{10}$
M$_{\odot}$). This is consistent with the result obtained by
\citet{zhgibo2012} that NGC 3115 has a relatively low number of LMXBs
per unit stellar mass compared with other galaxies studied by them
(NGC 3115 was included in their galaxy sample with only the first
three observations analyzed). Their fit with the normalization
decreased by 24\% is also shown in Figure~\ref{fig:lumdis_sum_subcxb}
(blue dashed line).

We note that we also found a high-break fit (Table~\ref{tbl:xlffit},
purple dot-dashed line in Figure~\ref{fig:lumdis_sum_subcxb})
that gives a higher break luminosity ($(1.6\pm0.1)\times10^{38}$
erg~s$^{-1}$) and a much steeper second slope ($>6.4$, 1-$\sigma$
lower limit) and has a C statistic only larger than the above
low-break fit by 1.1. The initial slope of the high-break fit
($1.13\pm0.08$) is slightly higher than that of the low-break
fit. 

\subsection{The XLF of LMXBs in the Field}
\label{sec:fieldxlf}
The incompleteness corrected and CXB contribution subtracted XLF above
$L_{\rm lim}$ for LMXBs in the field is plotted in the middle panels
in Figure~\ref{fig:lumdis_sum_subcxb}. The results of our fit with a
broken PL are given in Table~\ref{tbl:xlffit}. As for the XLF of all
LMXBs, we also find a low-break fit and a high-break fit to the XLF of
field LMXBs. The low-break fit has a C statistic higher than the
high-break fit by only 0.3. The parameters of both fits are very
similar to the corresponding fits to the XLF of all LMXBs
(Section~\ref{sec:xlfs}), probably due to relatively few GC LMXBs in
the total sample. Figure~\ref{fig:lumdis_sum_subcxb} shows
  both the low-break fit (green solid line) and the high-break fit
  (purple dot-dashed line) to the field LMXB XLF.

To check whether there is spatial variation of the XLF of field LMXBs,
we divided our study field into an inner ((0.046--0.2)$D_{\rm 25}$)
and an outer ((0.2--1.0)$D_{\rm 25}$) region (the boundary was chosen
to ensure enough statistics in both regions) and created two
corresponding XLFs.  They are shown in
Figure~\ref{fig:fieldlmxbxlf_sub}.  It is nontrivial to use the K-S
test to compare these two XLFs due to their different CXB and
incompleteness corrections. Therefore, we also fitted them with a
broken PL for comparison. The fitting results are shown in the figure
and given in Table~\ref{tbl:xlffit}. The XLF of field LMXBs in the
inner region seems to be flatter at low luminosities, and the break
seems to be at a lower luminosity than that in the outer region. We
see no clear degeneracy in the fit any more, but the fit to the XLF from
the inner region is similar to the low-break fit to the XLF from the
whole region, and the fit to the XLF from the outer region is similar to
the high-break fit to the XLF from the whole region, especially the
break and second slope. The best-fitting $\alpha_1$, $L_{\rm b}$, and
$\alpha_2$ of the XLFs in the inner and outer regions differ at the
1.5$\sigma$, 1.8$\sigma$, and 1.5$\sigma$ confidence levels,
respectively. Such differences are marginally significant, and we will
discuss the possible origin in Section~\ref{sec:dis}.

\subsection{The XLF of LMXBs  in GCs}
\label{sec:gcxlf}
The XLF for the LMXBs detected in {\it HST}/ACS GCs are shown in the
right panels in Figure~\ref{fig:lumdis_sum_subcxb}. It is
incompleteness corrected using the incompleteness function (green
dashed line) shown in Figure~\ref{fig:incompletefun}. The XLF of GC
LMXBs seems flatter than that of field LMXBs. The fit with a broken PL
is given in Table~\ref{tbl:xlffit}. We obtained
$\alpha_1=0.10^{+0.39}_{-0.62}$, $\alpha_2=1.6\pm0.2$, and $L_{\rm
  b}=11^{+2}_{-3}$, which are different from those from the field
LMXBs (the low-break fit) by 2.5$\sigma$, 1.6$\sigma$ and 2.3$\sigma$,
respectively. Our results agree with previous finding that the XLF of
the field LMXB is steeper than that of GC LMXBs
\citep{vogisi2009,kifabr2009,zhgivo2011}, though the difference is not
very significant in our case, owing to the relatively few GCs in our sample.

Previous studies of the XLF of GC LMXBs often just used sources within
the $D_{\rm 25}$ ellipse. We also obtained such an XLF for GC LMXBs,
and the fitting results with a broken PL are given in
Table~\ref{tbl:xlffit}. We obtained $\alpha_1$, $\alpha_2$, and
$L_{\rm b}$ different from those from the field LMXBs (the low-break
fit) by $1.5\sigma$, 1.5$\sigma$, and 1.1$\sigma$, respectively. Due
to fewer sources used, these differences are less significant than
those reported above using all {\it HST}/ACS GCs.

\subsection{Optical Properties of GC LMXBs}
\label{sec:gclmxbs}
Figure~\ref{fig:gcpro} plots the color-magnitude diagram of {\it
  HST}/ACS GCs. To indicate the presence of LMXBs and their luminosity
we circle the 37 GCs containing LMXBs, with the size of the circle
proportional to the logarithm of the LMXB luminosity, but for 13 GCs
whose $L_{\rm X}$ clusters within (0.8--1.4)$\times10^{37}$ erg
s$^{-1}$, they were enclosed with squares instead. We note that all
our GC LMXBs have $L_{\rm X}>1.13\times10^{36}$ erg s$^{-1}$. The
median ($g-z$) colors of GCs with LMXBs and without LMXBs are 1.23 and
1.06, respectively.  Based on the nonparametric Wilcoxon rank sum
test, the difference is at a significance level of 2.7$\sigma$.
Concentrating on luminous LMXBs with $L_{\rm X}>10^{37}$ erg s$^{-1}$
(24 in total), we found that the median ($g-z$) colors of GCs with
luminous LMXBs and without luminous LMXBs are 1.31 and 1.05,
respectively, corresponding to a Wilcoxon rank sum difference of
3.5$\sigma$.  If we follow \citet{jestro2014} and use $g-z=1.13$ mag
as the dividing line (dotted line in Figure~\ref{fig:gcpro}) between
the red/metal-rich and blue/metal-poor subpopulations, we find that 23
out of 169 (i.e., 13.6\%) red GCs contain LMXBs, while there are 14
out of 191 (i.e., 7.3\%) blue GCs containing LMXBs. Thus, the fraction
of red GCs hosting LMXBs is about twice of that of blue GCs hosting
LMXBs. Concentrating on luminous LMXBs with $L_{\rm X}>10^{37}$ erg
s$^{-1}$, we find 18 (i.e., 10.7\%) red GCs and 6 (i.e., 3.1\%) blue
GCs hosting luminous LMXBs. The former fraction is 3.5 times of the
latter, which is consistent with previous studies using a sample of
galaxies with limiting X-ray luminosity around $10^{37}$ erg s$^{-1}$
\citep{kumaze2007, sijosa2007}.

Figure~\ref{fig:gcpro} also shows that GCs hosting LMXBs tend to be
bright/massive, as found previously for many galaxies
\citep[][including NGC 3115, but using only the {\it Chandra}
observation in 2001]{sijosa2007,kumaze2007}. The median $M_z$ is
$-9.62$ for GCs with LMXBs and is $-8.23$ for GCs without LMXBs,
corresponding to a Wilcoxon rank sum difference of
6.1$\sigma$. Similar results can be obtained if we just focus on
luminous LMXBs with $L_{\rm X}>10^{37}$ erg s$^{-1}$, with the median
$M_z$ of $-9.48$ for GCs with luminous LMXBs and $-8.29$ for GCs
without luminous LMXBs (the Wilcoxon rank sum difference is
4.4$\sigma$). Separating the GC subpopulations and concentrating on
bright GCs with $M_z<-9.0$ mag, we find 16 out of 53 (i.e., 30\%) red
GCs and 10 out of 36 (i.e., 28\%) blue GCs hosting LMXBs. For even
brighter GCs with $M_z<-10$, we find 10 out of 14 (i.e., 71\%) red GCs
and 3 out of 8 (i.e., 38\%) blue GCs containing LMXBs. These fractions
are significantly higher than those obtained above for all GCs (i.e.,
13.6\% and 7.3\% for red and blue GCs, respectively).

\section{DISCUSSION}
\label{sec:dis}
\subsection{The Correction of CXB Contribution in XLFs}
We have obtained the XLFs of LMXBs in NGC 3115 down to $L_{\rm
  lim}\approx10^{36}$ erg s$^{-1}$, which has only been achieved for
one other old galaxy, i.e., Centaurus A \citep{vogisi2009}. We have
carried out careful corrections to the XLFs to account for the
incompleteness effects and CXB contribution. We found no large
discrepancy between the CXB density in our field and that estimated by
\citet{genala2008}, who used data from six large {\it Chandra}
surveys. Even assuming a possible 20\% enhancement of the CXB density
in our field (Section~\ref{sec:spatialdis}), we found no noticeable
effect on the XLFs. In some studies, the CXB contribution was taken
into account by directly excluding CXB sources identified from the
optical cross-correlation \citep[e.g.,][]{kifabr2009}. We did not show
the results using this method because our {\it HST} imaging is not
deep enough. However, we also tested this method by excluding the AGNs
that we identified in Paper I (about 50\% of the expected number) and
obtained XLFs very similar to the ones that we have shown. This is
mainly because of the high inclination of NGC 3115 so that the CXB
contribution is less significant for this galaxy than other typical
ones.

\subsection{Caveats on the XLF of Field LMXBs and Physical
  Implications}
Our XLF of field LMXBs can be fitted with parameters typically seen in
the literature for other old normal galaxies or the bulge of spiral
galaxies, with a possible break around $5.7\times10^{37}$ erg
s$^{-1}$. Such a break has been attributed to different mechanisms of
removal of orbital angular momentum \citep[magnetized stellar wind
versus gravitational wave emission,][]{poku2005} or different types of
donor stars \citep[giants versus main-sequence stars,][]{repoku2011}
in the high and low luminosities.

However, the interpretation of our XLF of field LMXBs is complicated
by the presence of a degenerate solution with a high break at about
$2\times10^{38}$ erg~s$^{-1}$. This degeneracy appears to be
associated with a spatial variation of the XLF, with the XLF in the
inner region ((0.046--0.2)$D_{\rm 25}$) being flatter and having a
lower break ($3\times10^{37}$ erg s$^{-1}$) than that in the outer
region ((0.2--1.0)$D_{\rm 25}$), which has a break around
$1.7\times10^{38}$ erg s$^{-1}$. Therefore, the XLF of field LMXBs in
the inner region is closer to the XLF of GC LMXBs.  One possible cause
for this is that our field LMXB sample in the inner region could
include some GC LMXBs that we cannot identify due to significant
incompleteness effects near the galaxy center that limit our ability
to detect GCs in the optical. In the outer region we found 25 LMXBs
from 219 GCs and 44.5 in the field (after excluding the CXB
contribution) above $L_{\rm lim}$, but only 2 GC LMXBs from 7 GCs and
43.0 field ones in the inner region. To have the same ratio of GC
LMXBs to field LMXBs, we would have missed 22 GC LMXBs in the inner
region. However, it is well known that the spatial distribution of GCs
is more extended than the stellar light, as can also be seen in
Figure~\ref{fig:spatialdist} for the outer region. The distribution of
GCs within (0.3--1.3)$D_{\rm 25}$, if fitted with a PL, is found to
follow $\sqrt{\alpha/R_{\rm 25}}$, which would indicate that in the
inner region there should be 93 GCs. Assuming the same detection rate
of LMXBs ($25/219=11.9\%$) as in the outer region, we would have
missed 6.6 GC LMXBs, given that we have detected 2 GC LMXBs and 2
candidates. Thus, the number of GC LMXBs that we missed in the inner
region is probably small, compared with the total number of sources
observed (43.0 after subtracting the CXB contribution), and their
effect on the XLF of field LMXBs in the inner region should be
insignificant.

Alternatively, the spatial variation of the XLF of field LMXBs might
be real and can be explained if the field LMXBs in the inner region
have a dynamical origin similar to GC LMXBs. There are two scenarios:
one is the dynamical formation of LMXBs in the dense stellar
environment near galaxy nuclei, and the other is the destruction of
GCs that drift toward the galaxy center due to mass segregation,
leaving behind the remnant LMXBs. The former was argued to be the
dominant mechanism to account for the high specific frequency of X-ray
sources, per unit stellar mass (following the $\rho_*^2$ dependence on
the stellar density), near the center ($<$$1\arcmin$) of M31 by
\citet{vogi2007a,vogi2007b}. However, we do not see increasing high
specific frequency of (non-GC) X-ray sources at the very center,
compared with the outer region (Figure~\ref{fig:spatialdist}). The
stellar density of M31 is around 30 \msun\ pc$^{-3}$ at $1\arcmin$
from the center \citep{vogi2007b}. Based on the stellar density model
by \citet{emdeba1999}, we expect that NGC 3115 reaches a similar
stellar density at $\alpha$$\sim$8$\arcsec$ in the major-axis
direction. Therefore the former mechanism is probably still not
significant in our inner region, which excludes the central
$a=10\arcsec$ elliptical region. There is large uncertainty in
estimating the level of the second mechanism. The specific frequency
of (non-GC) X-ray sources seems to peak around $\alpha/R_{\rm 25}\sim$
0.1--0.2. If some part of it is due to the second mechanism, the
remnant LMXBs should gain some momentum to reach this region during
the destruction of GCs, or the destruction should be able to occur
there.

Considering the possible contamination of the LMXBs from undetected or
disrupted GCs in the inner region, the XLF in the outer region is
probably a better representation of the characteristics of primordial
field LMXBs. Its break at about $1.7\times10^{38}$ erg s$^{-1}$
(Table~\ref{tbl:xlffit}) seems somewhat higher (at the 90\% confidence
level) than typical values of around $5\times10^{37}$ erg s$^{-1}$
reported in other studies that normally used most of the field (that
is, no differentiation between the inner and outer regions)
\citep{gi2004,vogisi2009,kifabr2009,zhgibo2012}. \citet{zhgibo2013}
obtained XLFs combining 20 early-type galaxies for the inner and outer
regions separately. Their inner and outer regions were defined as
(0.2--3)$r_{\rm e}$ and (4--10)$r_{\rm e}$, respectively, where
$r_{\rm e}$ is the $K_{\rm s}$-band half-light radius. Our inner and
outer regions for NGC 3115 are approximately (0.3--1.3)$r_{\rm e}$ and
(1.3--6.5)$r_{\rm e}$, respectively. Visually it appears that the
break of the XLF in their outer region is higher than the XLF in their
inner region (see their Figure~4). However, they did not carry out the
fit, and the significance of this variation is not clear. Moreover,
they did not exclude GC LMXBs, making it difficult to compare
directly. In the future, more galaxies should be used to investigate
the XLF of field LMXBs from the outer region to check whether the high
break that we observed in the XLF of field LMXBs in the outer region
of NGC 3115 is universal or due to statistical fluctuation. If it is
real, the best explanation is probably the Eddington limit of NS
LMXBs. In Paper I, we have shown that most of our bright LMXBs (above
several 10$^{36}$ erg~s$^{-1}$) are NS LMXBs in the soft state.

In summary, the flatter XLF of field LMXBs in the inner region
compared to that in the outer region is unlikely due to dynamically
formed LMXBs in the dense stellar environment near the galaxy nucleus,
but could be due to contamination of LMXBs from undetected and/or
(more likely) disrupted GCs in the inner region. The field LMXBs in
the outer region are more likely to be primordial. The break of their
XLF could be due to the Eddington limit of NS LMXBs, agreeing with our
finding in Paper I that most of our bright sources are NS LMXBs in the
soft state.

\subsection{Caveats on the XLF of GC LMXBs and Physical Implications}
Considering the large detection rate of LMXBs ($\sim$$71\%$,
Section~\ref{sec:gclmxbs}) in the most metal-rich and the most massive
GCs, some of these GCs probably in fact host multiple LMXBs which cannot
be resolved by {\it Chandra}. We follow the method of
\citet{sijosa2007} to study such source blending effects. They found
the dependence of the expected number $\lambda_{\rm t}$ (assuming
Poisson statistics) of LMXBs per GC on the GC properties to be:
\begin{equation}
\lambda_{\rm t}=A\left(\frac{M}{10^6\ 
    M_{\odot}}\right)^{1.237}10^{0.9(g-z)}\left(\frac{r_{{\rm h},M}}{\rm 1\ pc}\right)^{-2.22},
\label{eq:gcprob}
\end{equation}
where the GC mass is $M=1.45\times10^{-0.4(M_z-M_{z,M_{\odot}})}$
\msun\ ($M_{z,M_{\odot}}=4.512$) and the half-mass radius is
$r_{h,M}=r_{\rm h}\times10^{0.17[(g-z)-1.22]}$. To match our
observation of 37 GCs hosting LMXBs, the normalization, $A$, should be
at least 0.16, without taking into account the incompleteness
effects. The expected median $M_z$ is then $-9.36$, slightly fainter
than the observed value (i.e. $-9.62$, Section~\ref{sec:gclmxbs}).
Concentrating on the region of $M_z<-10.0$ and $(g-z)>1.13$ where the
LMXB detection rate is the highest, with 10 out of 14 GCs observed to
host LMXBs (Section~\ref{sec:gclmxbs}), the expected number of GCs
hosting LMXBs from Equation~\ref{eq:gcprob} with $A=0.16$ is 8.0, with
the number of GCs expected to host multiple LMXBs $N_{\rm
  multi}=3.5$. The total expected number of LMXBs is $N_{\rm
  LMXB}=13.5$. To match the observed value of 10 GCs hosting LMXBs,
$A$ should be 0.25, resulting in $N_{\rm multi}=5.8$ and $N_{\rm
  LMXB}=21.6$ in this region. A similar exercise for the rest of the
region (i.e., $M_z>-10.0$ or $(g-z)<1.13$), we find that $A=0.14$
could match the observed 27 GCs hosting LMXBs, with $N_{\rm
  multi}=3.4$ and $N_{\rm LMXB}=31.1$.

GCs hosting multiple LMXBs are expected to be more luminous and show
less long-term variability in X-rays \citep{kumaze2007}. In the dense
LMXB region of $M_z<-10.0$ and $(g-z)>1.13$, there are four with the
maximum 0.5--7 keV luminosity $>5\times10^{37}$ erg s$^{-1}$. They
have 0.5--7 keV long-term luminosity variability of $V=1.4$, 1.4, 2.1,
and 14.9. Among the field LMXBs within (0.046--1.0)$D_{25}$, there are
27 with maximum 0.5--7 keV luminosity $>5\times10^{37}$ erg
s$^{-1}$. Their median variability is 2.3, which is not significantly
larger than that found above for the GC LMXBs with high likelihood of
blending. Thus we cannot confirm any effect of blending on the
variability.

\begin{figure} 
\centering
\includegraphics[width=3.4in]{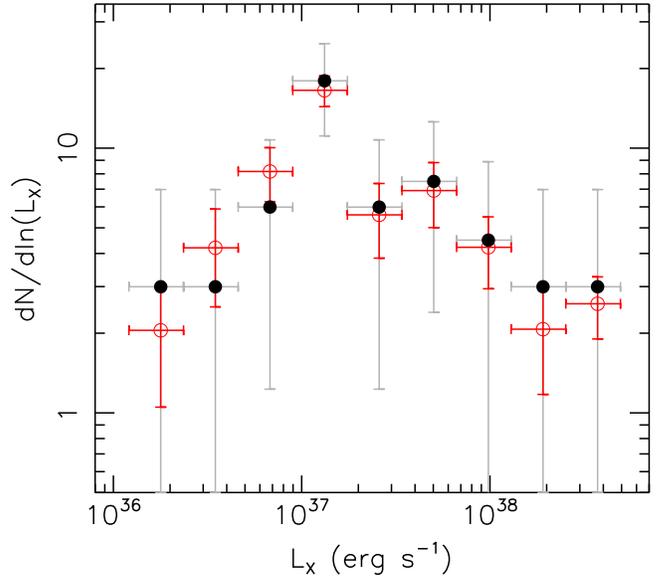}
\caption{The XLF of observed GC LMXBs (filled black circles) and the
  XLF using the mean of 1000 simulations (open red circles, with the
  standard deviation as the error bar). The simulations take into
  account the possibility that some GCs might host multiple LMXBs (see
  the text). The XLFs were not corrected for the incompleteness
  effects. \label{fig:gc_sim}}
\end{figure}

To investigate the source blending effects on the XLF, we carried out
Monte Carlo simulations using Equation~\ref{eq:gcprob} with $A=0.25$
if $M_z<-10.0$ and $(g-z)>1.13$ and $A=0.14$ elsewhere, as obtained
above. When a GC was simulated to host multiple LMXBs, we assumed them
to have equal luminosities. The results from 1000 simulations are
shown in Figure~\ref{fig:gc_sim}, where we plot the XLF using the mean
(open circles; the standard deviation is shown as the error bar) of
LMXBs in each luminosity bin from these simulations. The simulated
XLFs seem steeper than the observed one overall, as expected. However,
the difference is small, which is mainly due to two reasons. One is
that about one third of GC LMXBs are observed to cluster within a very
narrow luminosity range (0.8--1.4)$\times10^{37}$ erg s$^{-1}$, but
these GCs are widely spread around in the color-magnitude diagram,
thus producing no significant source blending effects. The second
reason is that in the area of parameter space where GCs preferentially
host LMXBs, both faint and luminous LMXBs were observed. That is, the
source blending effects are seen at different luminosity
levels. Therefore we conclude that the source blending effects on the
XLF are negligible, within the uncertainties of our data.

Instead of assuming equal weights to construct the XLF
(Section~\ref{sec:gcxlf}), we explored the option of weighting the
incompleteness function to reflect the tendency of LMXBs to be
detected in metal-rich and massive GCs. We recalculated the XLF of GC
LMXBs with the incompleteness function weighted by the probability of
hosting one or more LMXBs, based on Equation~\ref{eq:gcprob} again
with $A=0.25$ if $M_z<-10.0$ and $(g-z)>1.13$ and $A=0.14$ elsewhere
(the source blending effects cannot be taken into account
simultaneously though). We find that the XLF obtained in this way
shows no significant difference from that shown in
Section~\ref{sec:gcxlf}.

Therefore the observed paucity of faint GC LMXBs below the XLF break
$\sim$$10^{37}$ erg~s$^{-1}$ in NGC 3115 should be real. It could be
explained if there is a transition from persistent sources to
transients around this break \citep{vogisi2009,kifabr2009}. One main
class of GC LMXBs could be ultracompact X-ray binaries (UCXBs). These
are NSs accreting from white dwarf (WD) companions with very short
orbital periods ($P_{\rm orb}\lesssim1$ hr) and might be effectively
produced in GCs through direct stellar collisions between NSs and red
giants \citep{ve1987}.  \citet{bide2004} first suggested that UCXBs
with $P_{\rm orb}\sim8$--10 min could explain the XLF of GC LMXBs at
high luminosities. The second slope $\alpha_2=1.6\pm0.2$ that we
obtained in the broken PL fit to the XLF is consistent with their
prediction ($\alpha_2$=1.77). However, according to
\citet{ladukr2008}, for the transition from persistent to transient
behavior for a He-rich X-ray irradiated accretion disk to occur at
around $10^{37}$ erg~s$^{-1}$, systems with $P_{\rm orb}\gtrsim40$ min
are preferred, at least below the break luminosity.

It has been long-debated whether the entire population of LMXBs in
galaxies, including those in the field, was formed dynamically in GCs
\citep{whsaku2002,kumaze2002,kumaze2007,ir2005,ju2005,hubu2008}. Our
result that the XLFs of GC and field LMXBs appear to be different
agrees with previous findings
\citep[e.g.,][]{vogisi2009,kifabr2009,zhgivo2011}. The difference
indicates that they are formed through different channels that result
in different system configurations (orbital period, mass ratio, etc.)
and thus with different mass accretion rates. Therefore, our result
supports the idea that not all field LMXBs are formed dynamically in
GCs.

In summary, source blending should occur in GC LMXBs, but we
  do not expect it to significantly affect the XLF.  The observed
  paucity of faint GC LMXBs below the XLF break $\sim$$10^{37}$
  erg~s$^{-1}$, compared with field LMXBs, is likely real, and one
  explanation is that GC LMXBs are dominated by accreting neutron
  stars with white dwarf donors that show a transition from persistent
  sources at high luminosity to transients at low luminosity around
  this break. The different XLFs of GC and field LMXBs suggest
  that field LMXBs are not all formed dynamically in GCs.

\section{CONCLUSIONS}
\label{sec:con}
We have studied the XLF of LMXBs in the early-type galaxy NGC 3115,
using the the Megasecond {\it Chandra} XVP Observation of this
galaxy. Including three previous observations, we obtained a total
exposure of $\sim$1.1 Ms and reached a detection sensitivity of
$L_{\rm lim}\sim10^{36}$ erg s$^{-1}$, much lower than typically
achieved for other early-type galaxies ($L_{\rm lim}\sim10^{37}$ erg
s$^{-1}$) by {\it Chandra}. Our fit to the XLF of all LMXBs supports
the presence of the low-luminosity break of the XLF at around
$5.5\times10^{37}$ erg s$^{-1}$, with the differential PL slope of
$1.0\pm0.1$ and $2.2\pm0.4$ below and above the break, respectively,
as seen in many previous studies
\citep[e.g,][]{gi2004,vogisi2009,kifabr2009,zhgibo2012}. However, we
cannot exclude the possibility that the break is at around
$1.6\times10^{38}$ erg s$^{-1}$ and is sharp, with the differential PL
slope of $1.1\pm0.1$ and $>6.7$ below and above the break,
respectively.

We further created the XLFs for field and GC LMXBs separately. Due to
relative few GC LMXBs, the XLF of the field LMXBs is very similar to
the XLF of all (GC+field) LMXBs and still shows degeneracy in the
broken PL fit (one fit with a break at around $5.7\times10^{37}$ erg
s$^{-1}$ and the other fit with a sharp break at around
$2.0\times10^{38}$ erg s$^{-1}$). The field LMXB XLF seems to show
spatial variation, with the slopes and the break in the inner region
($(0.046$--$0.2)D_{\rm 25}$) being smaller than those in the outer
region ($(0.2$--$1.0)D_{\rm 25}$). This could be due to the
incompleteness effects of the optical GC detection in the inner region
and/or contamination of remnant LMXBs left behind from the destruction
of GCs that drift toward the galactic center due to mass
segregation. The XLF from the outer region is thus probably more close
to the XLF of primodial field LMXBs. It has a differential PL slope up
to a break at around $1.7\times10^{38}$ erg~s$^{-1}$, which is close
to the Eddington limit of NS LMXBs. The detection of spatial variation
explains the degeneracy in our fit to the XLF from the whole study
region.

The XLF of GC LMXBs overall is flatter than that of field LMXBs. Our
observation of the difference between the XLFs of GC and field LMXBs
casts doubt on the idea that all LMXBs in the galaxy are formed
dynamically in GCs. The break of the GC LMXB XLF is at around
$1.1\times10^{37}$ erg s$^{-1}$ and might be due to a transition from
persistent sources at high luminosity to transients at low luminosity,
which can be explained if GC LMXBs are dominated by accreting NSs with
WD donors.

As in previous studies, we found that metal-rich/red GCs are more
likely to host LMXBs than the metal-poor/blue ones, an effect that is
more significant for more luminous LMXBs, and that more massive GCs
are more likely to host LMXBs. Although source blending is likely to
occur, our simulations indicate that it should not significantly
affect the XLF.

While at the end of the preparation of this paper, \citet{lebeze2014}
also reported the study of three old normal galaxies. Their main goal
was to test the evolutionary model of LMXBs, but they also obtained
the XLFs of LMXBs in NGC 3115 using the same data presented here.  In
the Appendix, we briefly compared our study with theirs. We found no
large discrepancy between our results and theirs, if factors such as
the fitting degeneracy and the possible spatial variation are taken
into account.

The work is supported by {\it Chandra} XVP grant GO2-13104X. This
material is based upon work supported in part by the National Science
Foundation under Grants AST-1211995 and AST-1308124. This material is
based upon work supported in part by HST-GO-12759.02-A and
HST-GO-12759.12-A. GRS acknowledges support from an NSERC Discovery
Grant.

\appendix
\section{Comparison with Lehmer et al. 2014}
\label{sec:lehmer}
In their study of the evolutionary model of LMXBs, \citet{lebeze2014}
also presented the XLFs of LMXBs in three normal galaxies, including
NGC 3115. They used the same data presented here for NGC 3115, but
there are many differences between their analysis method and
ours. Lehmer et al. used the limiting significance level of $10^{-5}$
for {\it wavdetect} and kept sources with false binomial probability
less than 0.004 (see their Equation 1), while we adopted the limiting
significance level of $10^{-6}$ for {\it wavdetect} (Paper I) and used
sources above $L_{\rm lim}$, which results in using only sources with
the signal to noise ratio $\ge$2.9. Moreover, Lehmer et al. used
sources detected from the merged and individual observations (though
the false binomial probability was calculated exclusively from the
merged photometry), while we used sources detected from the merged
observation only. Therefore we expect that Lehmer et al. could detect
more real faint sources but also more spurious faint sources than we
did.

The region studied is also different. Lehmer et al. used an elliptical
region of a semi-major axis of 2.7$\arcmin$, a semi-minor axis of
1.1$\arcmin$ and a position angle of $45\degr$ (based on the $K$-band
galaxy emission), excluding a central circular region of radius
$20\arcsec$, but we use the slightly larger $D_{\rm 25}$ ellipse,
excluding the central $a=10\arcsec$ elliptical region. We excluded a
smaller central region because of our use of subpixel binning images
for source detection in the central region. Within their study region,
we have 90 sources above $L_{\rm lim}$, among which 17.4 are expected
to be CXB sources, while in our study region, we have 145 sources
above $L_{\rm lim}$, 26.5 of which are expected to be CXB sources. The
handling of the CXB contribution in the fits to the XLFs is also
different. Lehmer et al. excluded all AGNs that they could identify
from the {\it HST}/ACS imaging (they found 9 such sources, which is
about 50\% of the expected number), while we estimated the CXB
contribution following \citet{genala2008}.

We tried to check whether we can reproduce their XLF fitting results
(they also fitted the XLFs with a broken PL) based on the sources that
we detected but using their study region. We followed their technique
to exclude all AGNs that we could identify (10 from Paper I) from the
{\it HST}/ACS imaging instead of estimating the CXB contribution
following \citet{genala2008}. We found that we generally obtained
slightly lower values of the initial slope $\alpha_1$ by
(1--2)$\sigma$, most probably due to their inclusion of more very
faint sources than were in our sample. Specifically, in the fit to the
total XLF, they obtained $\alpha_1=1.5\pm0.1$, and we obtained
$\alpha_1=1.3\pm0.1$ if we chose a high-break solution with $L_{\rm
  b}=1.57_{-0.53}^{+0.24}\times10^{38}$ erg s$^{-1}$ similar to their
$L_{\rm b}=(1.76\pm0.02)\times10^{38}$ erg s$^{-1}$. We still see the
degenerate low-break ($L_{\rm b}=(5.0\pm0.2)\times10^{37}$ erg
s$^{-1}$) fit, which has a C statistic larger than that of the
high-break fit by only 0.4 (or by only 0.1 if the CXB contribution was
subtracted from modeling instead). For the field LMXB XLF, we
preferred a high-break fit, which was also adopted by Lehmer et al.
and is similar to our XLF of field LMXBs in the outer part of the
$D_{\rm 25}$ region (Section~\ref{sec:fieldxlf}). For the GC LMXB XLF,
they also obtained a high break luminosity $L_{\rm
  b}=(1.76\pm0.19)\times10^{38}$ erg s$^{-1}$, but we cannot constrain
it well ($L_{\rm b}=4.5_{-3.6}^{+14.2}\times10^{37}$ erg s$^{-1}$; we
have 22 GC LMXBs (including three candidates) above $L_{\rm lim}$ in
their region, while they have 25 GC LMXBs detected). Overall, we see
no large discrepancy between our results and theirs for LMXBs in their
study region. No large discrepancy is seen either between our results
for our study region and theirs, if factors such as the fitting
degeneracy and the possible spatial variation are taken into account.

\end{document}